\pdfoutput=1
\documentclass[aps,reprint,floatfix,superscriptaddress]{revtex4-1} 
\usepackage{amsmath,amsfonts,amssymb} %
\usepackage{mathtools} %
\usepackage{wasysym} %
\usepackage{bm} %
\usepackage[bbgreekl]{mathbbol} %
\usepackage{graphicx} %
\usepackage[dvipsnames,table]{xcolor} %
\usepackage{tabularx} %
\usepackage[acronym]{glossaries} %
\usepackage{braket} %
\usepackage[binary-units=true,detect-weight=true]{siunitx} %
\usepackage{fancyhdr} %
\usepackage{microtype} %
\usepackage{natmove}
\usepackage[caption=false]{subfig}
\usepackage[byname]{smartref} %
\usepackage[breaklinks, %
colorlinks, linkcolor=Blue, citecolor=Blue, urlcolor=Blue]{hyperref} %
\usepackage[version=3]{mhchem}
\AtBeginDocument{\usepackage{booktabs}} 

\def\MyTitle{Iterative Qubit Coupled Cluster approach with efficient
  screening of generators} %
\def\MyAuthora{Ilya G. Ryabinkin} %
\def\MyAuthorb{Robert A. Lang} %
\def\MyAuthorc{Scott N. Genin} %
\def\MyAuthord{Artur F. Izmaylov} %

\def\MySubject{Quantum computing, quantum chemistry} %

\hypersetup{ %
  pdftitle={\MyTitle{}}, %
  pdfauthor={\MyAuthora{}, \MyAuthorb{}, \MyAuthorc{}, and \MyAuthord{}}, %
  pdfsubject={\MySubject{}}, %
  pdfkeywords={quantum computing, NISQ devices, electronic structure methods} %
}

\DeclareMathOperator{\tr}{tr} %
\newcommand{\E}{\textrm{e}} %
\newcommand{\I}{\mathrm{i}\mkern1mu} %
\def\be{\begin{equation}} %
\def\ee{\end{equation}} %
\def\bea{\begin{eqnarray}} %
\def\eea{\end{eqnarray}} %

\DeclareSIUnit[number-unit-product = {\,}]\cal{cal}
\newcolumntype{Y}{>{\centering\arraybackslash}X}

\newacronym[longplural={degrees of freedom}, firstplural={degrees of
  freedom (DOF)}, plural={DOF}]{DOF}{DOF}{degree of freedom} %
\newacronym[longplural={equations of motion}, firstplural={equations
  of motion (EOM)}, plural={EOM}]{EOM}{EOM}{equation of motion} %

\newacronym{NISQ}{NISQ}{noisy intermediate-scale quantum}

\newacronym{JW}{JW}{Jordan--Wigner} %
\newacronym{BK}{BK}{Bravyi--Kitaev} %

\newacronym{QPE}{QPE}{quantum phase estimation} %
\newacronym{VQE}{VQE}{variational quantum eigensolver} %
\newacronym{QMF}{QMF}{qubit mean-field} %
\newacronym{QCC}{QCC}{qubit coupled cluster} %
\newacronym{iQCC}{iQCC}{iterative qubit coupled cluster} %
\newacronym{IQCC}{IQCC}{iterative qubit coupled cluster} %
\newacronym{PQA}{PQA}{parametrized quantum annealing} %
\newacronym{DIS}{DIS}{direct interaction set} %

\newacronym{CI}{CI}{configuration interaction} %
\newacronym{FCI}{FCI}{full configurational interaction} %
\newacronym{CASCI}{CASCI}{complete active space configurational
  interaction} %
\newacronym{MCSCF}{MCSCF}{multiconfigurational self-consistent
  field} %
\newacronym{CASSCF}{CASSCF}{complete active space self-consistent
  field} %
\newacronym{CC}{CC}{coupled cluster} %
\newacronym{UCC}{UCC}{unitary coupled cluster} %
\newacronym{UCCSD}{UCCSD}{unitary coupled cluster singles and
  doubles} %
\newacronym{CCSD}{CCSD}{coupled-cluster singles and doubles} %
\newacronym{CCSD-T}{CCSD(T)}{coupled-cluster singles and doubles and
  non-iterative triples} %
\newacronym{RHF}{RHF}{restricted Hartree--Fock} %
\newacronym{UHF}{UHF}{unrestricted Hartree--Fock} %
\newacronym{DMRG}{DMRG}{density-matrix renormalization group} %
\newacronym{DFT}{DFT}{density-functional theory} %

\newacronym{CAS}{CAS}{complete active space} %
\newacronym{PES}{PES}{potential energy surface} %
\newacronym{PEC}{PEC}{potential energy curve} %
\newacronym{MO}{MO}{molecular orbital} %

\newacronym{SQP}{SQP}{sequential quadratic programming} %
\newacronym{MMA}{MMA}{method of moving asymptotes} %

\begin{document}

\title{\MyTitle}

\author{\MyAuthora{}} %
\email{ilya.ryabinkin@otilumionics.com} %
\affiliation{OTI Lumionics Inc., 100 College St. \#351, Toronto,
  Ontario\, M5G 1L5, Canada} %

\author{\MyAuthorb{}} %
\affiliation{Department of Physical and Environmental Sciences,
  University of Toronto Scarborough, Toronto, Ontario, M1C 1A4,
  Canada; and Chemical Physics Theory Group, Department of Chemistry,
  University of Toronto, Toronto, Ontario, M5S 3H6, Canada}

\author{\MyAuthorc{}} %
\email{scott.genin@otilumionics.com} \affiliation{OTI Lumionics Inc.,
  100 College St. \#351, Toronto, Ontario\, M5G 1L5, Canada} %

\author{\MyAuthord{}} %
\email{artur.izmaylov@utoronto.ca} \affiliation{Department of Physical
  and Environmental Sciences, University of Toronto Scarborough,
  Toronto, Ontario, M1C 1A4, Canada; and Chemical Physics Theory
  Group, Department of Chemistry, University of Toronto, Toronto,
  Ontario, M5S 3H6, Canada}

\date{\today}

\begin{abstract}
  An iterative version of the \gls{QCC}
  method~[I.G.~Ryabinkin~\emph{et al.}, J.~Chem.~Theory~Comput.\
  \textbf{14}, 6317 (2019)] is proposed. The new method seeks to find
  ground electronic energies of molecules on \gls{NISQ} devices. Each
  iteration involves a canonical transformation of the Hamiltonian and
  employs constant-size quantum circuits at the expense of increasing
  the Hamiltonian size. We numerically studied the convergence of the
  method on ground-state calculations for \ce{LiH}, \ce{H2O}, and
  \ce{N2} molecules and found that the exact ground-state energies can
  be systematically approached only if the generators of the \gls{QCC}
  ansatz are sampled from a specific set of operators. We report an
  algorithm for constructing this set that scales linearly with the
  size of a Hamiltonian.
\end{abstract}

\glsresetall

\maketitle

\section{Introduction}

The advent of commercial quantum computers greatly stimulated a desire
to use them to solve practically relevant hard computational problems.
One such problem is the electronic structure
problem~\cite{Helgaker:2000}. Current and near-future quantum
computers are \gls{NISQ} devices~\cite{Preskill:2018/quant/79}, which
are restricted in the number of available qubits, in qubit
connectivity, and in the fidelity of single- and multi-qubit
entangling gates. Algorithms for such hardware need to minimize the
gate count and be able to withstand noise.

\Gls{VQE}~\cite{Peruzzo:2014/ncomm/4213, Wecker:2015/pra/042303} is
one such algorithmic framework. It engages both quantum and classical
computers in an iterative optimization of the system wave function
using the variational principle. The quantum computer constructs a
wavefunction guess
$\ket{\Psi(\boldsymbol \tau)} = {\hat U}(\boldsymbol \tau)\ket{0}$ as
a sequence of gates representing a parametrized unitary
${\hat U}(\boldsymbol \tau)$ acting on some initial qubit wave
function $\ket{0}$; $\boldsymbol \tau$ is a set of numerical
parameters. To obtain the expectation value of energy,
$E(\boldsymbol \tau) = \braket{\Psi(\boldsymbol \tau) | \hat H
  |\Psi(\boldsymbol \tau)}$, the quantum computer performs a series of
measurements, which involve $\ket{\Psi(\boldsymbol \tau)}$ and a qubit
Hamiltonian $\hat H$. $\hat H$ is derived from the second-quantized
form of the electronic Hamiltonian $\hat H_e$ of a problem using a
fermion-to-qubit transformation. The classical computer accepts the
energy estimate $E(\boldsymbol \tau)$ and provides an updated set of
parameters, $\boldsymbol \tau'$, to start the next cycle of the
algorithm.

The essential component of \gls{VQE} is a parametrized unitary
${\hat U}(\boldsymbol \tau)$ which defines a form (ansatz) of a wave
function and determines the accuracy of the method. The \gls{VQE} does
not specify it explicitly; the only practical constraint on
${\hat U}(\boldsymbol\tau)$ is its length when expressed in terms of
universal qubit gates.

One of the first ans\"atze explored for VQE was the \gls{UCCSD}
form~\cite{Peruzzo:2014/ncomm/4213,
  Mcclean:2016/njp/023023,OMalley:2016/prx/031007,
  Romero:2018/qct/014008,Hempel:2018/prx/031022,
  Nam:2019/ArXiv/1902.10171}. The \gls{UCC} parametrization has
several advantages: 1) it is systematically improvable due to its
clear fermionic excitation hierarchy, 2) it is
size-consistent~\cite{Crawford:2007-cc}, and 3) it is
variational~\cite{Taube:2006/ijqc/3393, Evangelista:2011/jcp/224102,
  Harsha:2018/jcp/044107}. Besides, \gls{UCC} is highly accurate
already at the \gls{UCCSD} level and rapidly convergent for molecules
near equilibrium configurations~\cite{Olsen:2000/jcp/7140,
  Larsen:2000/jcp/6677}. However, due to general non-commutativity of
involved operators, the \gls{UCC} form cannot be directly translated
into a sequence of quantum gates without an additional Trotter
approximation~\cite{Poulin:2015/qic/361, Romero:2018/qct/014008}.
Furthermore, fermionic excitation operators tend to produce redundant
terms in the qubit representation~\cite{Hempel:2018/prx/031022,
  Nam:2019/ArXiv/1902.10171}. This observation together with the
\gls{NISQ} hardware restrictions prompted a search for more efficient
\gls{UCC} forms~\cite{Lee:2019/jctc/311, Nam:2019/ArXiv/1902.10171}.

On the other hand, experimental simulations of small molecules carried
out on an existing \gls{NISQ} device put forward a
``hardware-efficient'' ansatz~\cite{Kandala:2017/nature/242,
  Barkoutsos:2018/pra/022322}, which is a regular, periodic sequence
of parametrized single-qubit and fixed-amplitude two-qubit
\emph{gates}. While closely matching the hardware requirements, this
ansatz poses a problem of slow convergence with the number of gates.
The latter creates an overhead for the classical computer because
high-dimensional global minimization with parametrized circuits scales
exponentially with the number of
dimensions~\cite{Mcclean:2018/nc/4812,Lee:2019/jctc/311}.

A hardware-oriented approach has stimulated an interest in
methods~\cite{Ryabinkin:2018/jctc/6317,
  Grimsley:2018/arXiv/1812.11173} that operate directly in the space
of multi-qubit operators:
\begin{equation}
  \label{eq:pauli_word_def}
  \hat P_k = \prod_{i} {\hat\sigma_i^{(k)}}, \ \sigma \in \{\hat
  1, \hat x, \hat y, \hat z\}. 
\end{equation}
One such method, the \gls{QCC}, was introduced in our early
work~\cite{Ryabinkin:2018/jctc/6317} and used the following ansatz:
\begin{align}
  \label{eq:qcc_wf}
  {\hat U}(\boldsymbol \tau)  = \prod_{k} \exp(-\I \tau_k \hat P_k/2). 
\end{align}
The total number of $\hat P_k$s grows exponentially with the number of
qubits. Therefore, to select the most relevant $\hat P_k$s, \gls{QCC}
uses a screening procedure. The screening is based on the energy
derivative with respect to the $\hat P_k$ amplitude, $\tau_k$, taken
at $\boldsymbol \tau = 0$. Thus, $\hat P_k$s with the largest energy
derivative magnitudes are included in Eq.~\eqref{eq:qcc_wf} first.
This ranking can be done efficiently on a classical computer. However,
the procedure's simplicity is outweighed by the exponential number of
operators that need to be tested, which limits the applicability of
the \gls{QCC} method.

In Ref.~\citenum{Grimsley:2018/arXiv/1812.11173}, the exponential
ranking problem of the \gls{QCC} method was avoided by limiting
$\hat P_k$s to a polynomial number of those that are produced by all
single and double fermionic excitations (an ``operator pool''). Since
this set cannot provide convergence to exact energy (otherwise the
\gls{UCCSD} method would be exact), an iterative scheme was employed:
at each iteration, $\hat P_k$s from the pool are ranked using partial
derivatives of energy with the wave function generated at the previous
iteration. Calculations of these derivatives are parallelizable, but
require a quantum device. Unfortunately, the iterative refinement of
the ansatz increases the size of the corresponding quantum circuit,
and eventually exhausts the capacity of an \gls{NISQ} device.
Moreover, convergence towards the exact energy is warranted only if
\emph{all} parameters in the ansatz at each iteration are fully
re-optimized. The computational cost of this optimization grows
exponentially with the number of parameters~\cite{Lee:2019/jctc/311}.

In this work we address two problems. First, how to construct and
characterize all operators that have significant energy derivatives
without the exponential screening procedure. Second, how to formulate
an iterative \gls{VQE}-type procedure that avoids expansion of a
quantum circuit by delegating additional work to the classical
computer and performing more measurements. The use of fixed-size
quantum circuits will enhance applicability of \gls{NISQ} devices for
solving the electronic structure problem.

The rest of the paper is organized as follows. First, after
introducing several prerequisites to the \gls{QCC} approach, we
formulate a new polynomially scaling generator screening procedure.
Second, we formulate an iterative \gls{QCC} scheme and estimate its
resource requirements. Third, approaches for reducing the iterative
\gls{QCC} resource requirements are discussed. Fourth, numerical
benchmarks for a few molecular systems (\ce{LiH}, \ce{H2O}, and
\ce{N2}) are presented.

\section{Theory}
\label{sec:glsiqcc-method}

\subsection{A few definitions}
\label{sec:electr-struct-he}

The \gls{QCC} method starts at the second-quantized electronic
Hamiltonian of a molecule:
\begin{equation}
  \label{eq:qe_ham}
  \hat H_e = \sum_{ij}^{N_\text{so}} h_{ij} {\hat a}^\dagger_i {\hat a}_j + \frac{1}{2}\sum_{ijkl}^{N_\text{so}}
  g_{ijkl} {\hat a}^\dagger_i {\hat a}^\dagger_j {\hat a}_l {\hat a}_k,
\end{equation}
where ${\hat a_i}^\dagger$ and ${\hat a_i}$ are fermion creation and
annihilation operators, while $h_{ij}$ and $g_{ijkl}$ are molecular
one- and two-electrons integrals, written in a spin-orbital basis. The
number of spin-orbitals, $N_\text{so}$, can be as large as
$2N_\text{AO}$, twice the number of orbitals in the atomic basis set
chosen for a molecule, but in the present work we consider
active-space Hamiltonians with $N_\text{so} < 2N_\text{AO}$. To obtain
an active-space Hamiltonian one has to prepare a basis of \glspl{MO},
typically by running the Hartree--Fock calculations and then
transforming one- and two-electron integrals to that basis. After
that, core (always occupied) and frozen virtual (always empty)
orbitals must be specified; the contribution of the latter to the
Hamiltonian~\eqref{eq:qe_ham} may be simply dropped, while the
contribution of the former must be re-calculated
explicitly~\cite{Helgaker:2000}. The resulting orbital count is:
$N_\text{so} = 2N_\text{AO} - 2N_\text{core} - 2N_\text{frozen}$,
where factors of 2 account for the doubling of the number of
spin-orbitals as compared to the number of spatial orbitals.



For VQE, the electronic Hamiltonian~\eqref{eq:qe_ham} is converted to
a qubit form by one of the fermion-to-qubit transformations,
\gls{JW}~\cite{Jordan:1928/zphys/631,AspuruGuzik:2005/sci/1704} or
\gls{BK}~\cite{Bravyi:2002/aph/210, Seeley:2012/jcp/224109,
  Tranter:2015/ijqc/1431, Setia:2017/ArXiv/1712.00446,
  Havlicek:2017/pra/032332}
\begin{equation}
  \label{eq:qubit_H}
  \hat H = \sum_{k=1}^{M} C_k \hat P_k,
\end{equation}
where $C_k$ are coefficients, and $\hat P_k$ are Pauli strings
[``words,'' see Eq.~\eqref{eq:pauli_word_def}]. The number of qubits
$n$ in $\hat H$ is equal to the number of spin-orbitals in the active
space, $n = N_\text{so}$. The total number of terms in $\hat H$, $M$
is $O(n^4)$, because the fermion-to-qubit transformations map each
term of $\hat H_e$ to the constant number of Pauli
words~\cite{Seeley:2012/jcp/224109}.

In QCC, an initial state $\ket{0}$ is parametrized as a direct product
of $n$ coherent
states~\cite{Radcliffe:1971/jpa/313,Arecchi:1972/pra/2211,Perelomov:1972,
  Lieb:1973/cmp/327}:
\begin{eqnarray}
  \label{eq:qmf_wf}
  \ket{\boldsymbol \Omega} & = & \prod_{j=1}^{n} \ket{\Omega_j}, \\
  \ket{\Omega_j} & = & \cos\left(\frac{\theta_j}{2}\right)\ket{\uparrow}_j + 
                       \E^{\I\phi_j}\sin\left(\frac{\theta_j}{2}\right)\ket{\downarrow}_j,
\end{eqnarray}
where $\ket{\uparrow}_j$ and $\ket{\downarrow}_j$ are eigenstates of
$\hat z_j$, and
$\boldsymbol{\Omega} = \{\theta_1, ..., \theta_n\}\cup \{\phi_1, ...,
\phi_n\}$ are the corresponding Bloch angles taken as variational
parameters. Such a parametrized product state is referred hereinafter
as the \gls{QMF} wave function~\cite{Ryabinkin:2018/jcp/214105}. The
total \gls{QCC} energy is
\begin{align}
  \label{eq:qcc_energy}
  E_\text{QCC}
  & = \min_{\boldsymbol \Omega, \boldsymbol \tau} \braket{\boldsymbol \Omega |
    {\hat U}(\boldsymbol\tau)^\dagger \hat H {\hat U}(\boldsymbol\tau)|\boldsymbol \Omega}.
\end{align}
To estimate the contribution of each $\hat P_k$ to
$\hat U(\boldsymbol\tau)$, one computes a modulus of the \gls{QCC}
energy derivative at $\boldsymbol \tau = 0$:
\begin{eqnarray}
  \label{eq:grad_eq}
  &&\left| \frac{\mathrm{d} E[\hat P_k] }{\mathrm{d}\tau_k} \Big|_{\boldsymbol\tau=0} \right| \nonumber
  \\
  & =& \left|\frac{\mathrm{d}}{\mathrm{d}\tau_k} \min_{\boldsymbol
       \Omega}\Braket{\boldsymbol \Omega | \E^{\I \tau_k \hat P_k/2} \hat H
       \E^{-\I \tau_k \hat P_k/2} | \boldsymbol \Omega}\Big|_{\tau_k =
       0}\right| \nonumber \\
  & =& \left|\Braket{\text{QMF} | -\frac{\I}{2} [\hat H,
       \hat P_k]|\text{QMF}}\right|\ge 0, \label{eq:nonzero_grad_cond}
\end{eqnarray}
where $\ket{\text{QMF}} = \ket{\boldsymbol \Omega_{\text{min}}}$ is
the \gls{QMF} wave function at the \gls{QMF} energy minimum.

\subsection{Efficient screening procedure}
\label{sec:impr-glsqcc-entanrgl}

Consider an arbitrary $\hat P_i$ that satisfies the gradient
condition~\eqref{eq:nonzero_grad_cond}. In the absence of external
magnetic fields, the electronic Hamiltonian in Eq.~\eqref{eq:qe_ham}
is real. This results in real coefficients of $\hat H$ and an even
number of $\hat y$ terms in $\hat P_k$s. Accounting for this, the
energy gradient for $\hat P_i$ can be rewritten as:
\begin{align}
  \label{energy_gradient_expanded}
  \frac{dE[\hat P_i]}{d\tau} = \sum_k C_k\,
  \mathrm{Im}\Braket{\boldsymbol\Omega_\text{min}| \hat P_k \hat P_i |
  \boldsymbol\Omega_\text{min}}.
\end{align}
For any $\ket{\boldsymbol\Omega_\text{min}}$, non-vanishing
contributions in Eq.~\eqref{energy_gradient_expanded} can be produced
only by $\hat P_k \hat P_i$ with purely imaginary matrix elements,
requiring $P_i$ to have odd powers of $\hat y$ terms.

Frequently, the optimized \gls{QMF} state
$\ket{\boldsymbol\Omega_\text{min}}$ is an eigenstate of all
$\{\hat z_i\}_{i=1}^n$ operators and, hence, any product of them:
\begin{equation}
  \label{perfect_QMF_assumption}
  \prod_{i} \hat z_i \ket{\boldsymbol\Omega_\text{min}} = \pm
  \ket{\boldsymbol\Omega_\text{min}}. 
\end{equation}
If the lowest-energy \gls{QMF} solution does not satisfy this
condition, one can define a ``purified'' mean-field state
$\ket{\Phi_0}$ that satisfies Eq.~\eqref{perfect_QMF_assumption} and
has the maximum overlap with the lowest-energy \gls{QMF} solution.
Thus, for further analysis, we will use $\ket{\Phi_0}$.

With a property of the independent-qubit reference
Eq.~\eqref{perfect_QMF_assumption}, the non-vanishing terms in
Eq.~\eqref{energy_gradient_expanded} are the products
$\hat P_k \hat P_i$ that contain only $\hat z$ operations, since
\begin{align}
  \label{eq:zero_cond}
  \braket{\Phi_0 | \hat f_j | \Phi_0} = 0,
\end{align}
where $\hat f_j \in \{\hat x_j, \hat y_j \}$ is a generalized
\textit{flip operator} acting on the $j^{\rm th}$ qubit. We denote the
\textit{flip indices} $F(\hat T)$ of a Pauli term $\hat P$ as
\begin{align}
  \label{eq:flip_indices}
  F(\hat P) = \{j : \hat f_j \in \hat P \}.
\end{align}

Using Eq.~\eqref{eq:zero_cond} we find that the only non-zero
contributions in $[\hat H, \hat P_i]$ for a given $\hat P_i$ are from
those $\hat P_k$s which have the same flip indices as $\hat P_i$.
This leads to partitioning of the original Hamiltonian as
\begin{align}
  \label{hamiltonian_sectors}
  \hat H & = \sum_k \hat S_k,
\end{align}
where
\begin{equation}
  \label{eq:S-def}
  \hat S_k = \sum_j C_j \hat P_j
\end{equation}
group the terms with the same flip indices,
\begin{align}
  F(\hat P_i) & = F(\hat P_j), \quad \forall (P_i, \hat P_j) \in \hat S_k. \label{sector_def}
\end{align}
Thus, Pauli words possessing the same flip indices introduce an
equivalence relation on the set of Hamiltonian terms. As a result, all
the generators $\hat P_i$ with even $\hat y$ parity or with
$F(\hat P_i)\ne F(\hat S_k),\forall k$ in
Eq.~\eqref{hamiltonian_sectors} have zero energy gradients.
Furthermore, any two generators $\hat P_i$ and $\hat P_j$ with the
same $\hat y$ parity and $F(\hat P_i) = F(\hat P_j)$ will have
identical gradients up to a sign, and hence the same gradient
magnitudes. Therefore, any generators obtained by replacements of
$\hat 1$ with $\hat z$ operators or permutations of $\hat x$ and
$\hat y$ that conserve $\hat y$ parity have the same gradient
magnitude. This leads to a set of $2^{n-1}$ generators that are
characterized by the same absolute energy gradients.

The number of equivalence classes [terms in
Eq.~\eqref{hamiltonian_sectors}] is bound from above by the total
number of terms in $\hat H$, $M = O(n^4)$. Thus, the set of all
operators that satisfy the gradient
condition~\eqref{eq:nonzero_grad_cond} has the size $O(M 2^{n-1})$
with a partitioning into $O(M)$ groups. We will refer to this set as
the \gls{DIS}.

A representative operator from the \gls{DIS} can be constructed as
follows. First, partition the Hamiltonian as in
Eq.~\eqref{hamiltonian_sectors} by grouping its terms according to
their flip indices. Second, for a given $\hat S_k$ take a product of
$\hat x$ operators for all but one index from $F(\hat S_k)$,
Eq.~\eqref{eq:flip_indices}, and multiply it by a single $\hat y$
operator with the remaining index. The resulting Pauli word $\hat P_k$
has the odd (1) number of $\hat y$ operators and is characterized by
the same flip set as $\hat S_k$. Third, compute the energy gradient by
Eq.~\eqref{eq:grad_eq} by taking $\hat S_k$ as $\hat H$ and $\hat P_k$
as a candidate; a modulus of the resulting value will characterize the
gradient group corresponding to $\hat S_k$. Finally, repeat these
steps for each $\hat S_k$ to find all representatives and their
gradients, thus obtaining the full description of \gls{DIS}. Since the
number of $\hat S_k$s is bound by the polynomial number of terms in
$\hat H$ the number of different gradient groups is also polynomial.
Due to symmetries encoded in the Hamiltonian coefficients $C_j$, some
of the representatives can have gradients close to zero.

\subsection{The \gls{iQCC}}
\label{sec:general-scheme}

By rewriting the \gls{QCC} energy expression~\eqref{eq:qcc_energy} as
\begin{equation}
  \label{eq:eqcc_alt}
  E_\text{QCC} = \min_{\boldsymbol \tau} \left\{ \min_{\boldsymbol \Omega} \braket{\boldsymbol
      \Omega| \hat{H}_{d}(\boldsymbol \tau) |\boldsymbol \Omega} \right\},
\end{equation}
one demonstrates that the \gls{QCC} energy is the minimum of the
\gls{QMF} minima for a canonically transformed (``dressed'')
Hamiltonian,
\begin{equation}
  \label{eq:UHU}
  \hat{H}_{d}(\boldsymbol \tau) = U^\dagger(\boldsymbol \tau) \hat H
  {\hat U}(\boldsymbol \tau),
\end{equation}
parametrized by the set of amplitudes $\boldsymbol \tau$.
$\hat{H}_d(\boldsymbol \tau)$ can be evaluated recursively as
\begin{align}
  \label{eq:qcc_func_single_tau}
  \hat{H}_{d}^{(k)}(\tau_k, \dots, \tau_1) = {}
  & \E^{\I \tau_k \hat P_k/2}\,
    \hat H_{d}^{(k-1)}(\tau_{k-1}, \dots, \tau_1) \, \E^{-\I \tau_k \hat
    P_k/2} \nonumber \\ 
  = {}
  & \hat{H}_{d}^{(k-1)} + \sin \tau_k
    \left(-\frac{\I}{2}\left[ \hat{H}_{d}^{(k-1)}, \hat
    P_k \right]\right) \nonumber \\
  & + \frac{1}{2}\left(1 - \cos \tau_k\right) \left(\hat P_k
    \hat{H}_{d}^{(k-1)} \hat P_k - \hat{H}_{d}^{(k-1)}\right) ,
\end{align}
where $k = 1, \dots, N_g$ and $\hat{H}_{d}^{(0)} = \hat H$. This
procedure produces $3^{N_g}$ distinct operator terms and exposes the
exponential complexity of the \gls{QCC} form for a classical computer.
However, as shown in Appendix~\ref{sec:comp-scal-dress}, if amplitudes
$\boldsymbol \tau$ are \emph{fixed}, the complexity of the dressing
Hamiltonian by Eq.~\eqref{eq:UHU} is $\sim M(3/2)^{N_g}$.

This observation suggests an iterative reformulation of the \gls{QCC}
procedure. Instead of a single-step optimization of $N_g > 1$
amplitudes, one can use multiple steps and optimize $N_g$ amplitudes
\emph{sequentially}. The number of operators introduced at each step
is a constant that can be as low as 1, which means that a quantum
circuit of a \emph{fixed size} can be used at each iteration. However,
this iterative formulation does not guarantee the convergence to the
exact answer. We did not find rigorous conditions when such
convergence was possible and resorted to numerical experiments (see
Sec.~\ref{sec:numerical-examples}).

The \gls{iQCC} algorithm is summarized below. The number of steps,
$N_\text{steps}$, and the number of generators, $N_g \ge 1$, which
will be used at each step, are parameters of the scheme. The
initialization step is the \gls{QMF} energy minimization to determine
$E^{(0)}_\text{QCC} = E_\text{QMF}$ and the initial set of Bloch
angles, $\boldsymbol\Omega^{(0)}$. The \gls{iQCC} loop is:
\begin{enumerate}
\item Run a generator sampling algorithm using the current Hamiltonian
  $\hat H_d^{(k-1)}$ and Bloch angles from a previous iteration (to
  construct the mean-field reference state,
  $\ket{\boldsymbol\Omega^{(k-1)}}$) to identify $N_g$ generators with
  the largest absolute gradients in a given pool. If the highest
  gradient is lower than a threshold, terminate the loop.

\item Minimize the \gls{QCC} energy, Eq.~\eqref{eq:qcc_energy} with
  respect to $N_g$ amplitudes and $2n$ Bloch angles starting from a
  random guess. If this search with a small (usually ~10) number of
  guesses fails to locate the solution with lower energy than on a
  previous iteration, perform an additional minimization using
  $\boldsymbol \tau = 0$ and Bloch angles from the previous iteration
  as a guess. The last attempt is \emph{guaranteed} to lower energy
  because the chosen generators have non-zero energy gradients by
  construction. The random-search stage is introduced to prevent
  sticking in local minima and saddle points. You may also terminate
  the procedure here if the energy difference between the current and
  previous iterations is below a threshold.

\item Replace the current Hamiltonian by its dressed version
  calculated by Eq.~\eqref{eq:qcc_func_single_tau} using the $N_g$
  amplitudes optimized at the current iteration.

\item Compress the resulting Hamiltonian using the techniques from
  Sec.~\ref{sec:compr-interm-hamilt} (optional).

\item If the number of steps exceeds $N_\text{steps}$, exit. Otherwise
  start a new iteration.
\end{enumerate}
The \gls{QCC} energy at exit is the final result: the ground-state
energy estimate for the Hamiltonian $\hat H$.

\subsection{Compression of intermediate Hamiltonians}
\label{sec:compr-interm-hamilt}

Since the size of the Hamiltonians increases in the course of the
iterations, it is natural to seek a method to ``compress'' them in
such a way as to guarantee that their ground-state energies differ
less than a desired accuracy $\epsilon$,
\begin{equation}
  \label{eq:spec_dist}
  |E_0(\hat{H}_d^{(k)}) - E_0(\hat{H}_{c}^{(k)})| \le \epsilon.
\end{equation}
One can set, for example, $\epsilon = \SI{1}{\milli\hartree}$, which
is better than the so-called ``chemical accuracy,''
\SI{1}{\kilo\cal\per\mol}.

The compression procedure that we propose is based on the Weyl's
spectral perturbation theorem~\cite{Weyl:1912/mathann/441,
  Bhatia:1996}:
\begin{equation}
  \label{eq:weyl_est}
  \max_{j} |\lambda_j^{\downarrow}(\hat H) -
  \lambda_j^{\downarrow}(\hat{H}_c)| \le \|\hat H - \hat{H}_c\| \le
  \|\hat H - \hat{H}_c\|_\mathrm{F}, 
\end{equation}
where $\{\lambda_j^{\downarrow}\}$ are eigenvalues of the
corresponding operator arranged in decreasing order,
$\|\hat A\| = \sup_{\|\Psi\| = 1} \|\hat A\Psi\|$ is the operator
norm, which is for a normal (diagonalizable) operator equal to
$\max_j\{\lambda_j\}$, and
$\| \hat A \|_\mathrm{F} = \sqrt{\tr{({\hat A}^\dagger \hat A)}}$ is
the Frobenius norm of $\hat A$~\footnote{To align our consideration
  with the Weyl's theorem, one should use the negate of the real
  Hamiltonian, $-\hat H$.}. The Frobenius norm of the qubit
Hamiltonian~\eqref{eq:qubit_H} is easy to evaluate:
\begin{equation}
  \label{eq:H_frob}
  \|\hat H \|_\mathrm{F} = 2^{n/2} \sqrt{ \sum_j |C_j|^2},
\end{equation}
since $\mathrm{Tr}{({\hat P}_i^\dagger \hat P_j)} = 2^n\delta_{ij}$,
where $\delta_{ij}$ is the Kronecker symbol. Thus, if all $|C_j|$ are
sorted in descending order, we can define an approximate (compressed)
Hamiltonian as
\begin{equation}
  \label{eq:compressed_H}
  \hat{H}_c = \sum_{j=1}^{J} C_j \hat P_j,
\end{equation}
where $J$ satisfes
\begin{equation}
  \label{eq:cutoff_eq}
  \sqrt{\sum_{j = J+1} |C_j|^2} \le \frac{\epsilon}{2^{n/2}}.
\end{equation}
At each iteration of the \gls{iQCC} method one can replace the dressed
Hamiltonian $\hat{H}_d^{(k)}$ with its compressed version,
$\hat{H}_c^{(k)}$, to use it as a starting operator for the next
iteration. According to inequality~\eqref{eq:weyl_est}, this will
change the spectrum by no more than $\epsilon$.

The suggested compression procedure is well-suited for use with the
\gls{iQCC} method. As we established in
Appendix~\ref{sec:comp-scal-dress}, the main reason for growing the
dressed Hamiltonians is the commutator term in
Eq.~\eqref{eq:qcc_func_single_tau}. However, its average value on the
\gls{QMF} wave function is precisely the value of the gradient
contribution of the corresponding generator $\hat P$, which is
systematically reduced by the \gls{iQCC} procedure. Thus, after
the initial rapid growth in size of the intermediate Hamiltonians, one
could expect progressively stronger compression when the commutator
contributions start systematically falling below the compression
threshold $\epsilon$. We verify these expectations numerically in
Sec.~\ref{sec:numerical-examples}.
\begin{table*}
  \centering
  \caption{Electronic structure calculations details and parameters of
    the second-quantized and qubit Hamiltonians for molecules used in
    the study}
  \begin{minipage}{\textwidth}
    \begin{tabularx}{1.0\textwidth}{>{}XYYY}
      \toprule
      Property                         &        \multicolumn{3}{c}{Molecule}                                                     \\ \cmidrule{2-4}
                                       & \ce{LiH}                         & \ce{H2O}\footnotemark[1]                           & \ce{N2} \\ \midrule
      Molecular configuration          & $d(\ce{LiH}) = \SI{1.7}{\angstrom}$ & $d(\ce{OH}) = \num{1.25}\text{ and }\SI{2.35}{\angstrom}$, $\angle \ce{HOH} = \SI{107.6}{\degree}$ & $d(\ce{NN}) = \SI{2.118}{\bohr} \approx \SI{1.12}{\angstrom} $  \\
      Assumed symmetry\footnotemark[2] & $C_{2v}$                          & $C_{2v}$                                                           & $D_{2h}$ \\
      Atomic basis set\footnotemark[3] & STO-3G                           & 6-31G                                                              & cc-pVDZ \\
      \Acrfull{MO} set                 &                       \multicolumn{3}{c}{Hartree--Fock (canonical)}                                             \\
      Total number of \glspl{MO}       & 6                                & 13                                                                 & 28      \\
      \Acrfull{CAS}                        &  2e/3orb ($2a_1$, $3a_1$, $4a_1$) & 4e/4orb ($1b_1$, $3a_1$, $4a_1$, $2b_1$)                           & 10e/8orbs (full valence) \\[1ex]
      Fermion-to-qubit mapping         & parity\footnotemark[4]           & \acrlong{BK}                                                       & parity\footnotemark[4] \\
      Spin-orbital grouping            &        \multicolumn{3}{c}{same spin: first all $\alpha$ then $\beta$} \\
      Number of qubits\footnotemark[5] & 4                                & 6                                                                  & 14      \\
      Length of the qubit $\hat H$     & 100                              & 165                                                                & 825     \\
      \bottomrule
    \end{tabularx}
    \footnotetext[1]{A similar setup was used in
      Ref.~\citenum{Abrams:2005/cpl/284}.} %
    \footnotetext[2]{The full symmetry groups for diatomics \ce{LiH}
      and \ce{N2} are $C_{\infty v}$ and $D_{\infty h}$, respectively,
      but the maximal Abelian subgroups with all-real irreducible
      representations were chosen instead as required by the
      \textsc{ALDET} module of the \textsc{GAMESS}
      program~\cite{gamessus, gamessus-2}.} %
    \footnotetext[3]{From the Basis Set Exchange
      library~\cite{emsl-2}.} %
    \footnotetext[4]{Described in
      Ref.~\citenum{Nielsen:2005/scholar_text}.} %
    \footnotetext[5]{After qubit reduction; see
      Sec.~\ref{sec:electr-struct-calc}.} %
  \end{minipage}
  \label{tab:op_prop}
\end{table*}

\section{Numerical examples}
\label{sec:numerical-examples}

\subsection{Electronic structure calculations and generation of qubit
  Hamiltonians}
\label{sec:electr-struct-calc}

Molecular configurations for all species used in this study, as well
as the atomic basis sets and active spaces, are listed in
Table~\ref{tab:op_prop}. One- and two-electron integrals for
Hamiltonians Eq.~\eqref{eq:qe_ham} in the Hartree--Fock \gls{MO}
basis, were computed by a locally modified version [Feb.~14, 2018
(R1)] of the \textsc{gamess} electronic structure
package~\cite{gamessus,gamessus-2}. Qubit Hamiltonians were derived
from their fermionic counterparts by applying the appropriate
fermion-to-qubit transformation (see Table~\ref{tab:op_prop}). In all
cases, the qubit Hamiltonians had stationary
qubits~\cite{Bravyi:2017/ArXiv/1701.08213}: namely, qubit operators at
positions $N_\text{so}$ and $N_\text{so}/2$, enter the Hamiltonian as
$\hat z$ or $\hat 1$. Thus, $z$-projections of these qubits are
constant (stationary), and the corresponding operators can be replaced
with their eigenvalues ($\pm 1$) to define qubit-reduced operators.
This procedure has been applied to all Hamiltonians, and final qubit
counts are listed in Table~\ref{tab:op_prop}. Eigenvalues of $\hat z$
operators for stationary qubits were chosen so as to ensure that the
ground states of the full and reduced Hamiltonians are the same. The
qubit reduction procedure has also been applied to operators other
than a Hamiltonian; see Appendix~\ref{sec:qubit-reduct-proc}.

\subsection{\ce{LiH} and \ce{H2O} near equilibrium geometry}
\label{sec:small-molecules-near}

For small-qubit problems like \ce{LiH} and \ce{H2O}, increasing the
size of intermediate Hamiltonians poses no difficulties, and no
mitigation techniques are necessary. Moreover, rapid convergence of
energy in this weak-correlation regime allowed us to gauge the ability
of the \gls{iQCC} method to approach the exact energy. We tested first
the most challenging case, $N_g =1$, in which only one amplitude is
optimized at each iteration. This regime highlights the importance of
choosing appropriate generators for the \gls{QCC} ansatz.

We compare three operator pools: all two-qubit Pauli operators, the
operators from spin-adapted single and double fermionic excitation
operators transformed to a qubit space (like in
Ref.~\citenum{Grimsley:2018/arXiv/1812.11173}), and the \gls{DIS}. In
any case, a single top-gradient [according to Eq.~\eqref{eq:grad_eq}]
generator of entanglement was chosen for the \gls{QCC}
ansatz~\eqref{eq:qcc_wf} at each iteration.

\begin{figure}
  \centering %
  \includegraphics[width=0.9\linewidth]{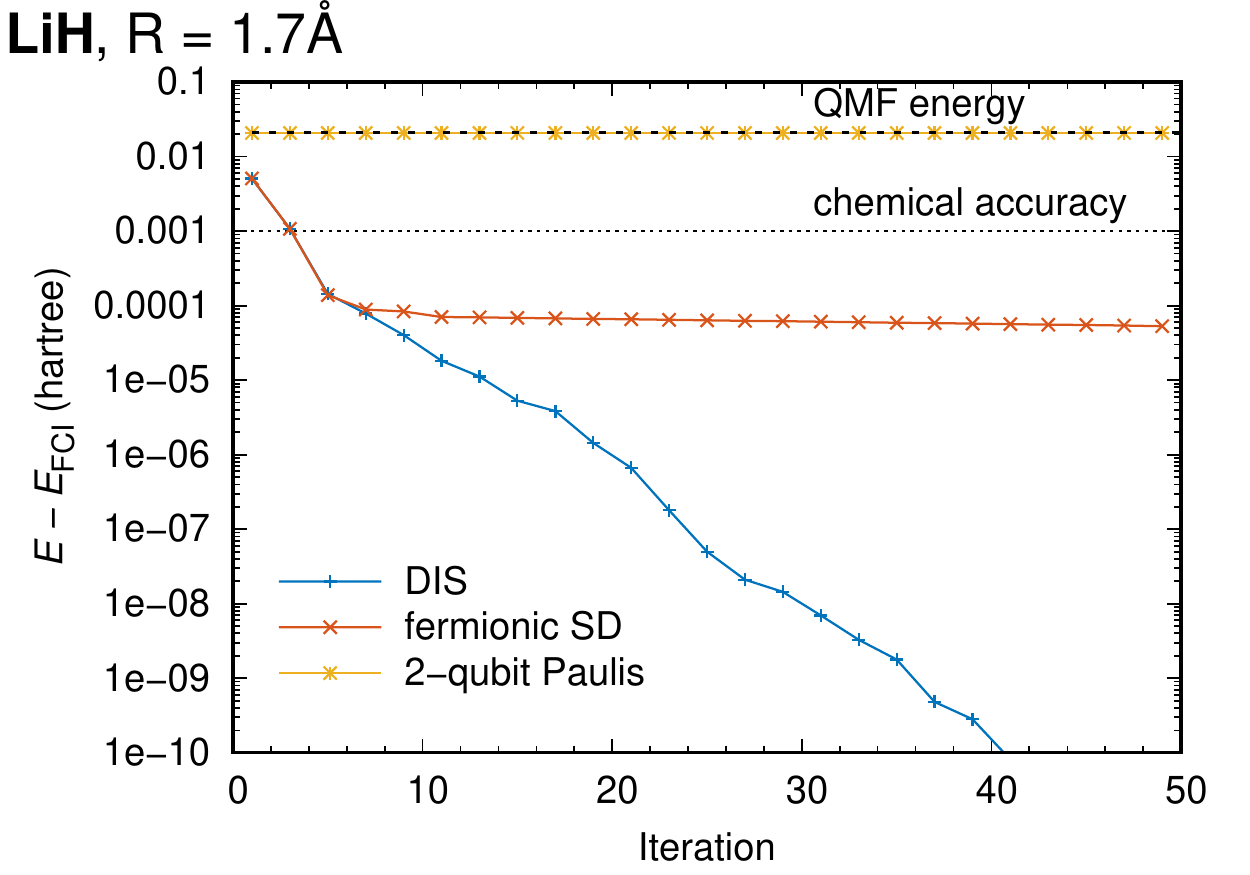}
  \includegraphics[width=0.9\linewidth]{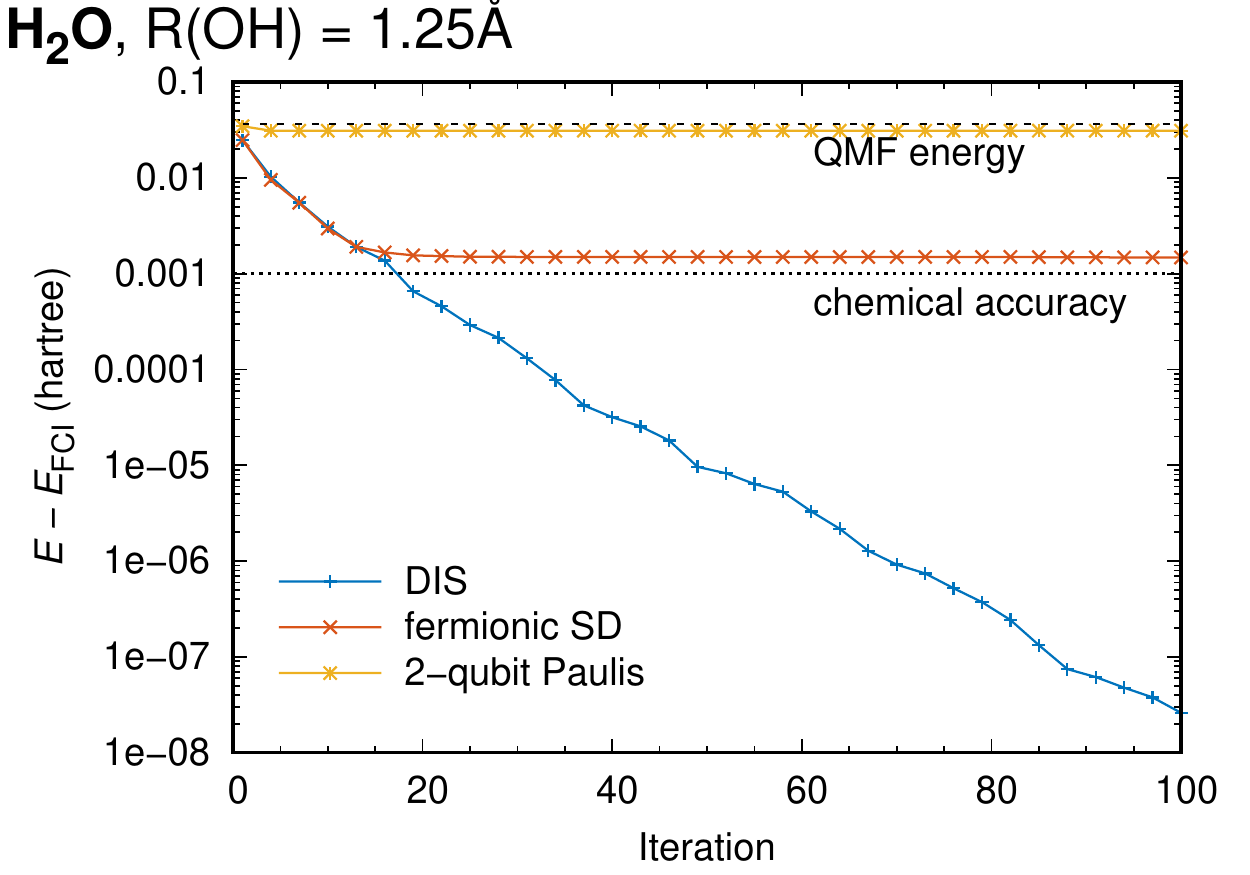}
  \caption{Convergence of a single-generator ($N_g =1$) \gls{iQCC}
    procedure for various choices of generator pools: ``fermionic SD''
    is a pool of Pauli operators that emerge after a fermion-to-qubit
    transformation of one- and two-body excitation operators, whereas
    the ``2-qubit Pauli'' is the pool of all two-qubit Pauli words.}
  \label{fig:1c_conv}
\end{figure}

Convergence of the \gls{iQCC} procedure with different operator pools
is shown in Fig.~\ref{fig:1c_conv}.
Energies for non-DIS pools saturate before reaching the exact energy,
while \gls{DIS} provides systematic and almost geometric convergence
towards the exact energy. We conjecture that the success of the
\gls{DIS} is due to its additional variational flexibility as it grows
together with the growth of intermediate Hamiltonians.

\subsubsection*{Adding multiple generators}
\label{sec:mult-entangl-case}

A faster converging setup for the \gls{iQCC} method is a scheme with
the addition of $N_g > 1$ generators at each step to fully exploit the
capabilities of an \gls{NISQ} device. However, in this case, a problem
of degeneracy appears. When generators are subjected to the
condition~\eqref{eq:nonzero_grad_cond}, some of them have identical
gradient magnitudes and, hence, cannot be ordered. This is a
consequence of the structure of the \gls{DIS}, which is revealed in
Sec. ~\ref{sec:impr-glsqcc-entanrgl}, but to the extent in which the
other operator pools overlap with the \gls{DIS}, it is relevant to
them as well. The simplest possible strategy in this situation is a
stochastic sampling, in which we draw a single random representative
from $N_g$ highest-gradient groups. We believe that this procedure
reduces the probability of being trapped in local minima.

\begin{figure}
  \centering %
  \includegraphics[width=1.0\linewidth]{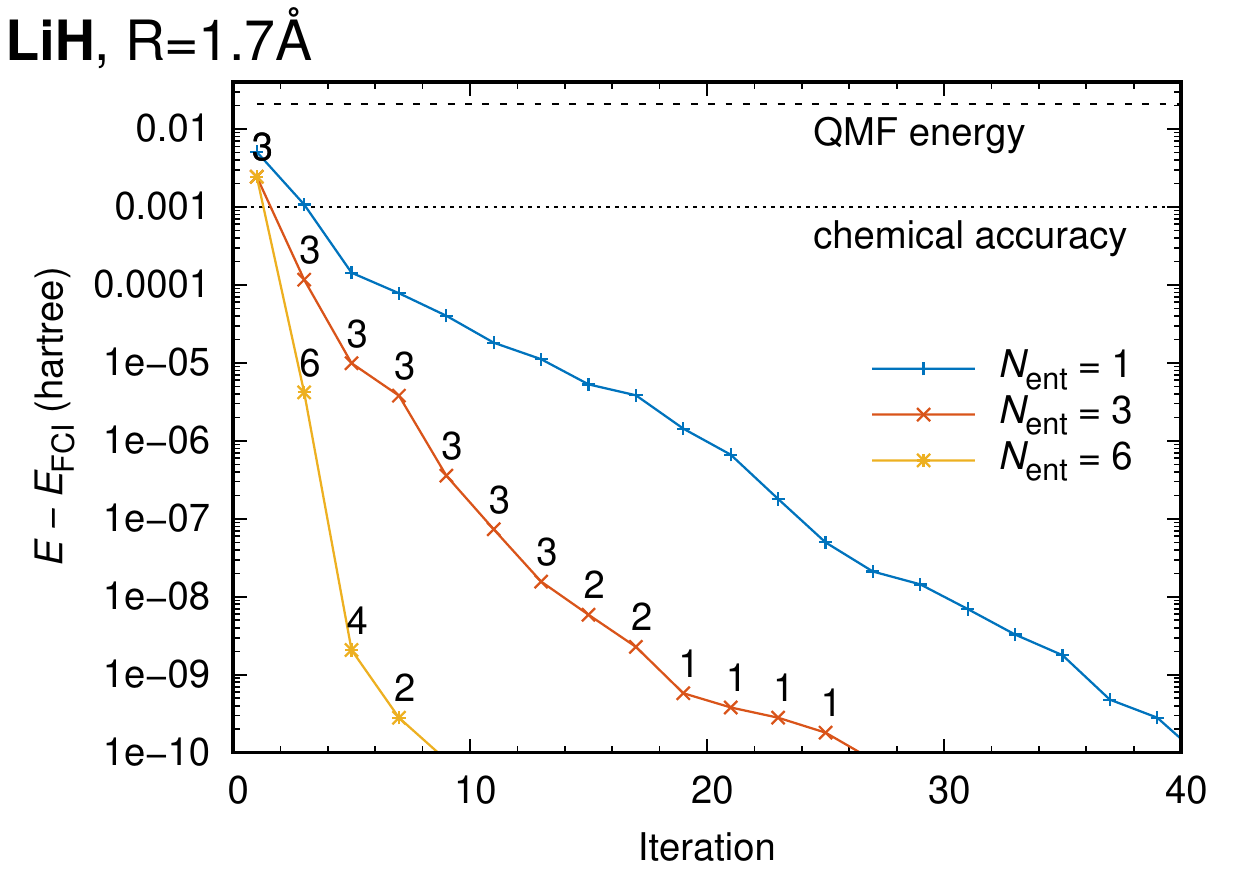}
  \caption{Convergence of the \gls{iQCC} procedure for different
    numbers of generators $N_g$ from the \gls{DIS}. The numbers beside
    each point for $N_g > 1$ are the numbers of generators chosen by
    the iterative procedure. They can be less than $N_g$ because the
    number of gradient groups is less than $N_g$.}
  \label{fig:multiple_entanglers_conv}
\end{figure}

With multiple generators a faster convergence is observed; see
Fig.~\ref{fig:multiple_entanglers_conv}. Only 5 iterations are needed
for $N_g = 6$ to bring the energy closer than $\SI{1e-8}{\hartree}$ to
exact for the \ce{LiH} molecule. A similar trend was observed for the
\ce{H2O} molecule (not shown).

Sometimes grouping found less than $N_g$ distinct gradient groups. In
this case $N_g$ is dynamically adjusted to match the number of groups;
these numbers are also shown in
Fig.~\ref{fig:multiple_entanglers_conv}.
We do not expect this problem to be important for larger-qubit
problems, since the number of gradient groups will be much larger than
$N_g$.

\begin{figure}
  \centering %
  \includegraphics[width=1.0\linewidth]{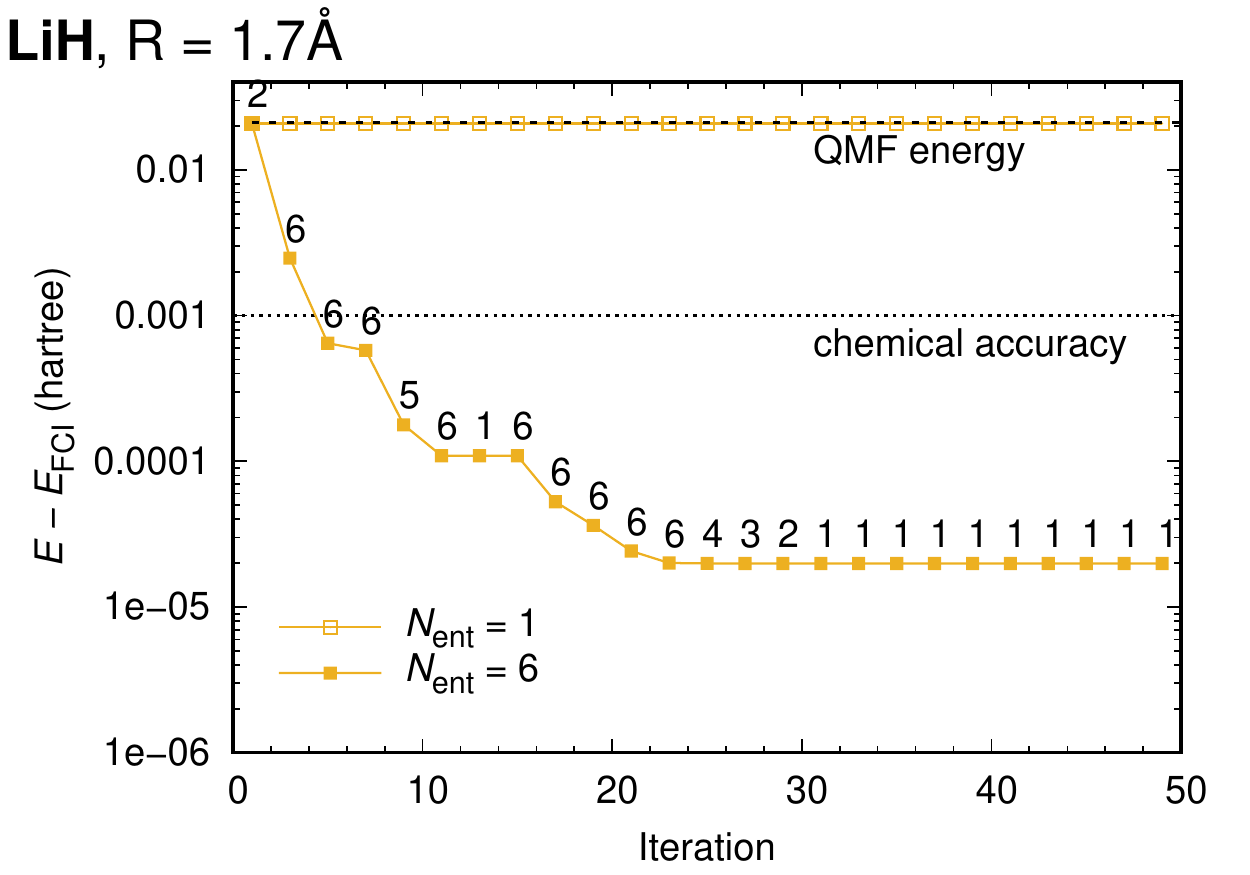}
  \caption{Same as in Fig.~\ref{fig:multiple_entanglers_conv} but with
    two-qubit Pauli words as the operator pool.}
  \label{fig:me_fr2}
\end{figure}

Larger \gls{QCC} ansatz can also mitigate the issue of premature
convergence when the limited operator pools are used. For example, we
found (see Fig.~\ref{fig:me_fr2}) that with $N_g = 6$ the \gls{iQCC}
procedure with sampling from the pool of two-qubit Pauli operators
converges to a much lower (albeit not to the exact) energy than the
same procedure with $N_g = 1$ (\emph{cf.}\ Fig.~\ref{fig:1c_conv}).
Thus, with a more sophisticated \gls{QCC} form, and a more capable
\gls{NISQ} device the use of limited operator pools may be justified.

\subsection{Stretched \ce{H2O}}
\label{sec:stretch-water-molec}

To assess the \gls{iQCC} method to treat strong correlation, we
consider a stretched \ce{H2O} molecule with
$\ce{R(O-H)} = \SI{2.35}{\angstrom} \approx
2.5\,R_e$~\footnote{$R_e = \SI{0.95}{\angstrom}$, is the \ce{O-H}
  distance in the water molecule at equilibrium geometry}. To obtain
the singlet \gls{QMF} solution, we used the penalty
method~\cite{Ryabinkin:2018/arXiv/1812.09812}, where a spin-penalized
Hamiltonian
\begin{equation}
  \label{eq:H_S2penalty}
  \hat H + \frac{\mu}{2} \hat S^2, \ \mu >0
\end{equation}
for sufficiently large $\mu$ ($\ge$\SI{0.25}{\hartree}) is used.
Noticeable deviations from the spin purity in the course of iterations
are tolerable because the exact ground state is a singlet, and, as
long as the convergence is established, spin deviations vanish.

\begin{figure}
  \centering %
  \includegraphics[width=0.9\linewidth]{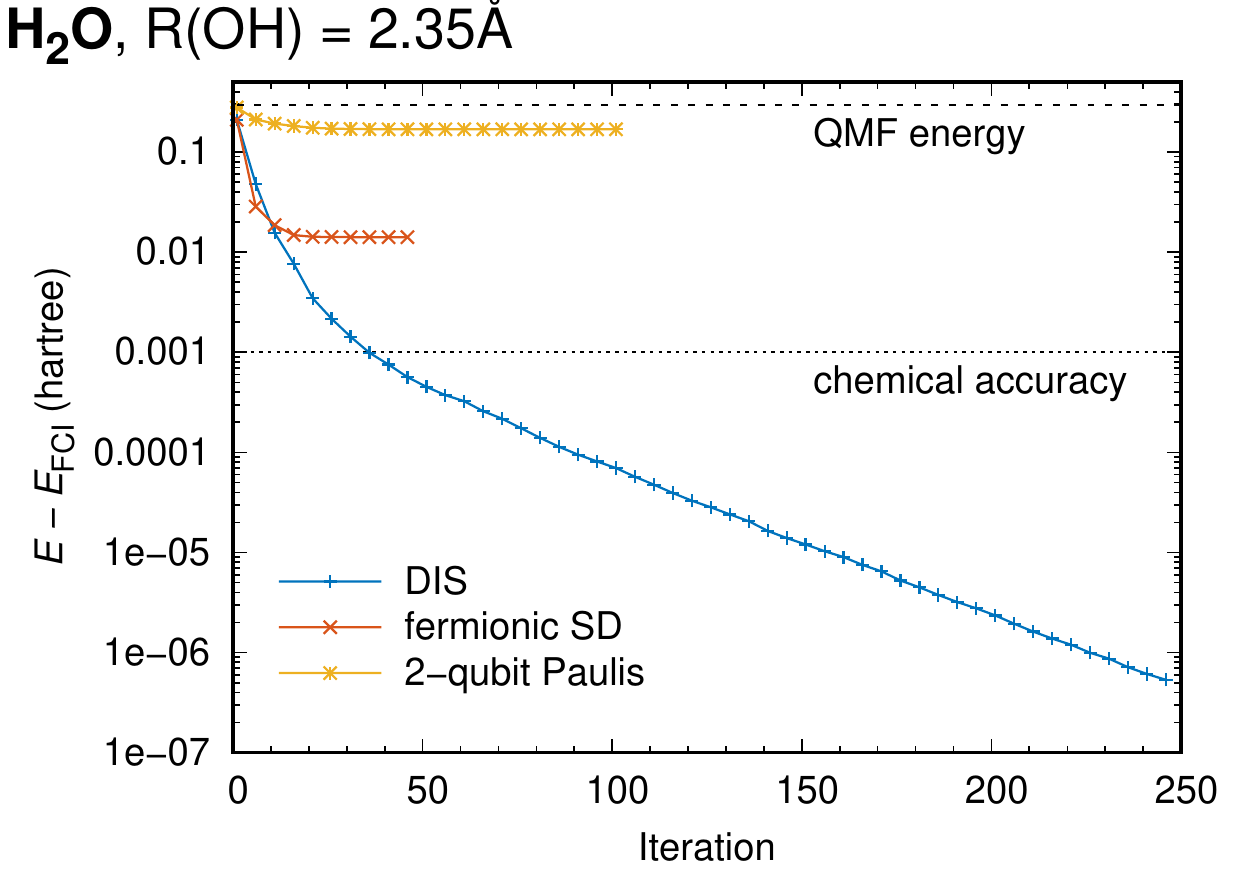}
  \caption{Same as in Fig. \ref{fig:1c_conv}. For the fermionic and
    two-qubit Pauli words pools convergence was reached to a given
    tolerance before the required number of iterations was performed.}
  \label{fig:1c_conv_strong_corr}
\end{figure}

The convergence of the \gls{iQCC} procedure with different operator
pools is shown in Fig.~\ref{fig:1c_conv_strong_corr}.
Compared to the weak-correlation case (Fig.~\ref{fig:1c_conv}) the
convergence is slower. Three times more iterations are necessary to
reach similar accuracy with the \gls{DIS}. Moreover, the initial onset
of geometric convergence lasts longer, and the convergence curve is
non-linear in the logarithmic scale for the first $\sim 70$
iterations. The other pools exhibit the same behaviour as in the
weak-correlation case. They level off before reaching the exact
energy. This behaviour is due to a sequential optimization of
amplitudes in the ansatz. The amplitudes from earlier iterations
cannot be adjusted later. Surprisingly, the number of terms in dressed
Hamiltonians at intermediate iterations are comparable between
different pools, suggesting that the \gls{DIS} pool surpasses its
competitors by providing more efficient dressing.

\subsection{Compression and extrapolation for large-qubit problems}
\label{sec:compr-interm-hamilt-1}

\begin{figure}
  \centering %
  \includegraphics[width=1.0\linewidth]{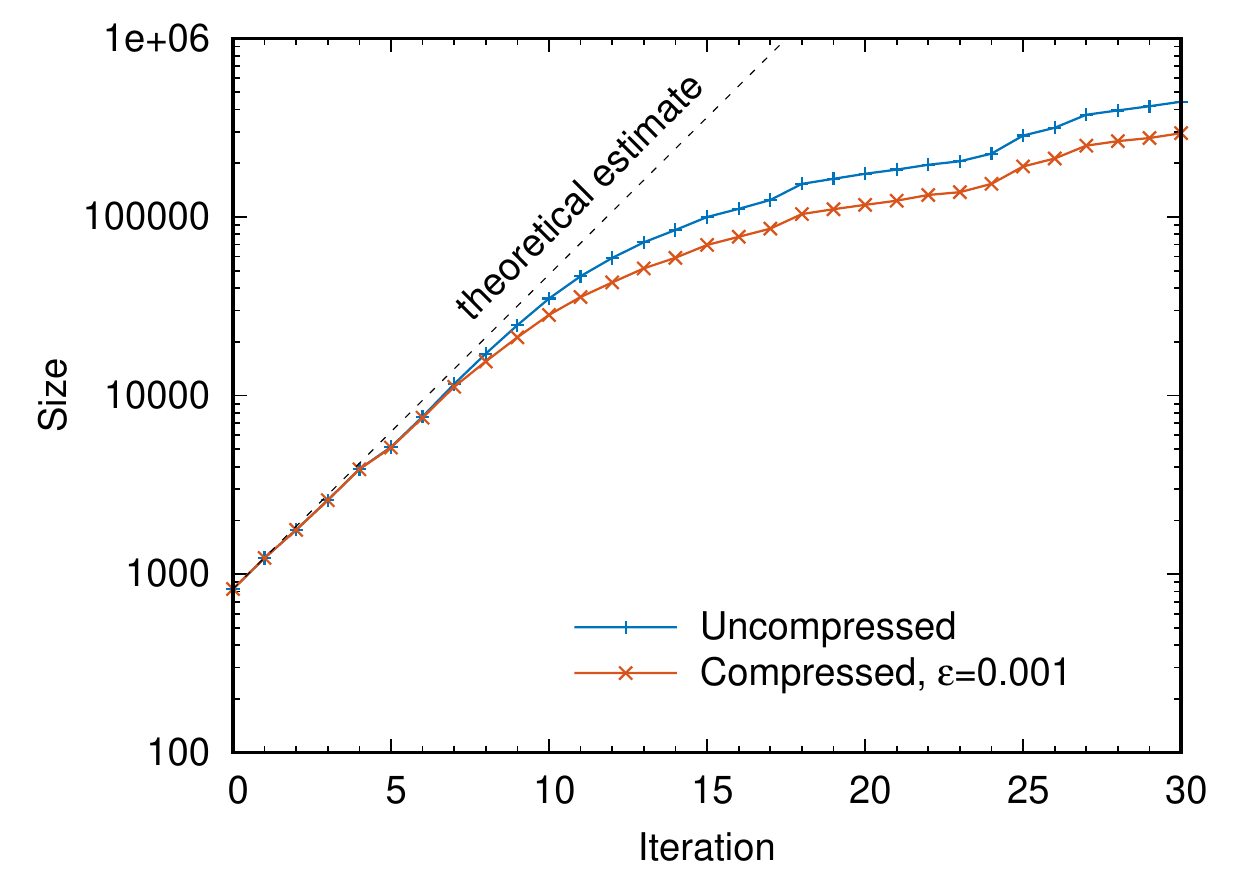}
  \caption{The number of terms of uncompressed and compressed
    Hamiltonians at each iteration of the \gls{iQCC} procedure for the
    \ce{N2} molecule with $N_g = 1$. Theoretical estimate assumes a
    $M \left(\frac{3}{2}\right)^{\text{Iteration}}$ law derived in
    Appendix~\ref{sec:comp-scal-dress}, where $M = 825$ is the size
    of the initial Hamiltonian (see Table~\ref{tab:op_prop}).}
  \label{fig:compressed_size}
\end{figure}

We consider a 14-qubit problem for \ce{N2} near the equilibrium
geometry (\emph{i.e.}\ in the weak-correlation regime) as a prototype
for ``large-scale'' problems. In the tensor-product basis the
Hamiltonian matrix has the size $2^{14} \times 2^{14}$, and we expect
a rapid growth of the size of intermediate dressed Hamiltonians. This
is confirmed by the results in Fig.~\ref{fig:compressed_size}. After
30 iterations, the uncompressed canonically transformed Hamiltonian
has $\sim\num{5e5}$ terms.
The onset stage, in which the growth follows the trend predicted in
Sec.~\ref{sec:general-scheme},
$M \left(\frac{3}{2}\right)^{\text{Iteration}}$, continues for the
first $~10$ iterations. After that the proliferation of terms slows
down.

Compression allows one to have $\sim 1.5$ times fewer terms with a
negligible impact on energies (Fig.~\ref{fig:compressed_size}). The
\gls{QCC} estimates at iterations differ by less than
\SI{3e-6}{\hartree} without apparent accumulation of errors. Thus, the
spectral compression can be applied with even more aggressive
thresholds due to a rather conservative estimate of ground-state
disturbance.

\begin{figure}
  \centering %
  \includegraphics[width=1.0\linewidth]{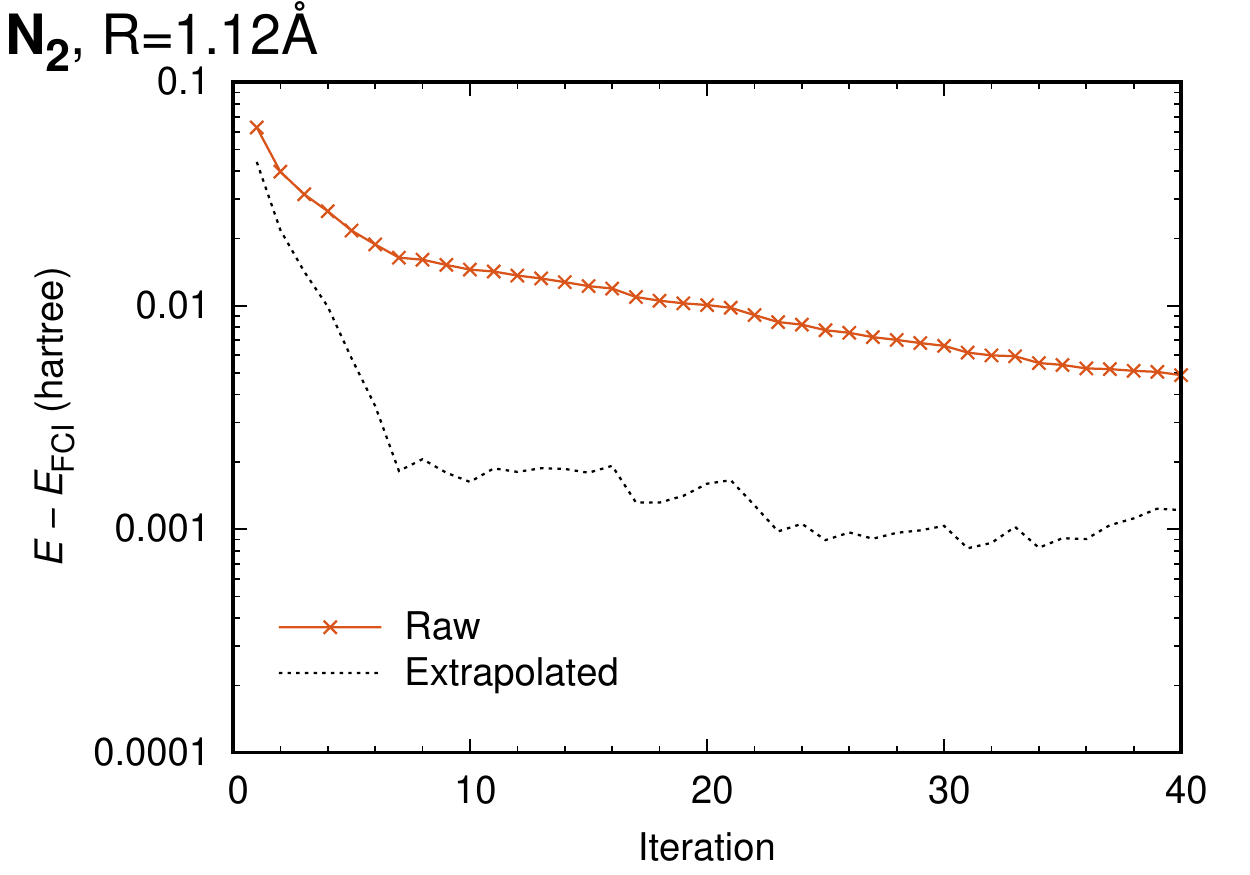}
  \caption{Convergence of raw and extrapolated by
    Eq.~\eqref{eq:exact_estimane} energy estimates for the \gls{QCC}
    procedure with $N_g =1$.}
  \label{fig:N2_conv}
\end{figure}

Having fewer terms in the Hamiltonian noticeably improves efficiency
of the \gls{iQCC} procedure. However, even after 40 iterations the
deviation from the exact energy is more than \SI{1}{\milli\hartree}
(see Fig.~\ref{fig:N2_conv}), so that comparable accuracy in
\emph{relative} energies cannot be guaranteed. Noticing a regular
(geometric) convergence pattern of the \gls{iQCC} energies, we suggest
using \emph{extrapolation} to improve final estimates. This technique
gained some popularity, for example, in the \gls{DMRG}
calculations~\cite{Marti:2010/mp/501}.

As evident from Fig.~\ref{fig:N2_conv}, the convergence is close to
geometric (linear in the logarithmic scale) after the first $\sim 10$
iterations.
If one assumes perfect geometric convergence, then
\begin{equation}
  \label{eq:exp_ideal}
  \log{\left(E_\text{QCC}^{(k)} - E_\text{exact}\right)} = ak + b.
\end{equation}
Of course, the exact energy $E_\text{exact}$ is not known. However, by
exponentiating, subtracting, and taking the logarithm again one can
obtain
\begin{equation}
  \label{eq:exp_sequential}
  \log{\left(E_\text{QCC}^{(k-1)} - E_\text{QCC}^{(k)}\right)} = a'k + b',
\end{equation}
where
\begin{align}
  \label{eq:fit_parameters}
  a & = a', \\
  b & = b' - \log{\left(10^{-k'} - 1\right)}
\end{align}
with an obvious requirement $k' < 0$. $a'$ and $b'$ can be determined
via the least-square fitting of Eq.~\eqref{eq:exp_sequential}, and the
exact energy estimate at each iteration may be computed as
\begin{equation}
  \label{eq:exact_estimane}
  E_\text{exact} = E_\text{QCC}^{(k)} - 10^{(ak + b)}.
\end{equation}
We dropped the first 10 values from \gls{iQCC} iterations and used the
remaining ones to fit Eq.~\eqref{eq:exp_sequential} and to calculate
the extrapolated values by Eq.~\eqref{eq:exact_estimane}. For the last
10 iterations, the extrapolated value was around
\SI{-109.0340}{\hartree}, which is \SI{1}{\milli\hartree} above the
exact one \SI{-109.0350}{\hartree}, a 5-fold increase in precision as
compared to the raw values. Thus, extrapolation may be useful for
large-scale \gls{iQCC} iterations.

\section{Conclusions}
\label{sec:conclusions}

We have presented and assessed a new technique for the ground-state
electronic structure calculations dedicated specifically for use with
\gls{NISQ} devices -- the \acrfull{iQCC} method. It is an iterative
modification of our previous technique, the \gls{QCC}
method~\cite{Ryabinkin:2018/jctc/6317}, with a few important
amendments.

First, a new selection procedure for generators of the \gls{QCC}
ansatz~\eqref{eq:qcc_wf} is proposed. It has computational complexity,
which is linear in the number of terms ($M$) of the Hamiltonian, a
significant improvement over the previous, exponentially difficult
procedure. Apart from that, the new procedure unveiled a non-trivial
structure of the set of operators satisfying the \gls{QCC} selection
condition, Eq.~\eqref{eq:nonzero_grad_cond}. This \acrlong{DIS} is a
union of $O(M)$ groups, each of which contains precisely $2^{n-1}$
operators with identical gradient magnitudes.

Secondly, our numerical experiments indicate that the \gls{iQCC}
procedure converges to the exact energy even with a fixed-size
\gls{QCC} ansatz when the \gls{DIS} is used as a pool of generators.
Compared to competitive approaches, such as the ADAPT-VQE
method~\cite{Grimsley:2018/arXiv/1812.11173}, the \gls{QCC} method
requires neither holding of the whole ansatz on a quantum device nor
cumbersome full re-optimization of all its parameters, thus enhancing
practical utility of \gls{NISQ} devices.

It must be admitted that these advantages are not free of cost. The
canonical transformation [Eq.~\eqref{eq:UHU}] increases the size of
the Hamiltonian up to a factor of $3/2$ at each step, which means an
increased load on a classical computer and more measurements for a
quantum one. Recently, multiple proposals for improving the
measurement efficiency have appeared~\cite{Jena:2019/arXiv/1907.07859,
  Verteletskyi:2019/arXiv/1907.03358,
  Izmaylov:2019/arXiv/1907.09040,Huggins:2019/arXiv/1907.13117,
  Crawford:2019/arXiv/1908.06942, Zhao:2019/arXiv/1908.08067}. It
would be interesting to apply them to the \gls{iQCC} intermediate
Hamiltonians. To partially compensate an increased load on a
\emph{classical} computer, we introduced two techniques, compression
and extrapolation. The first is a rigorous scheme for removing
``unimportant'' terms from the canonically-transformed Hamiltonians,
while the second is inspired by numerical behaviour of the \gls{iQCC}
energy estimates. Both techniques improve the efficiency of
large-scale \gls{iQCC} calculations.

We believe, therefore, that the \gls{iQCC} method is a viable
alternative to other approaches for electronic structure calculations
on \gls{NISQ} devices with the unique feature of employing fixed-size
quantum circuits, though its systematic convergence is shown only
numerically.

\appendix

\section{Computation scaling of the canonical transformation,
  Eq.~\eqref{eq:UHU}}
\label{sec:comp-scal-dress}

Consider dressing of the original Hamiltonian
$\hat{H}_{d}^{(0)} = \hat H$, which has $M = O(n^4)$ terms, with
$\hat U(\tau) = \exp(-\I \tau\hat P/2)$, for a fixed value of $\tau$. For
$k = 1$, Eq.~\eqref{eq:qcc_func_single_tau} contains three distinct
operator terms, $\hat H$, $[\hat H, \hat P]$, and
$\hat P \hat H \hat P$ multiplied by numerical coefficients. First, we
notice that the operator $\hat P \hat H \hat P$ has the \emph{same}
operator terms as $\hat H$. This is a simple consequence of the
anti-commutation relations for Pauli elementary operators:
$\hat\sigma' \hat\sigma + \hat\sigma \hat\sigma' = 0$
($\hat\sigma \ne \hat\sigma'$). Multiplying this relation by
$\hat\sigma$ from the left and using the involutory property,
$\hat \sigma^2 = 1$, we write:
\begin{equation}
  \label{eq:pauli_comm}
  \hat \sigma \hat \sigma' \hat \sigma =
  \begin{cases}
    -{\hat \sigma'}, & \hat  \sigma' \ne \hat \sigma \\
    \sigma, & \sigma' = \hat \sigma.
  \end{cases}
\end{equation}
Thus, every term of $\hat P \hat H \hat P$ has either the same or the
opposite sign compared to that in $\hat H$. That is, the sum of
$\hat H$ and $\hat P \hat H \hat P$ with an arbitrary numerical
coefficient has no more terms than $\hat H$ itself. Therefore, new
algebraically independent terms may only come from the commutator term
$[\hat H, \hat P]$.

An arbitrary Pauli word commutes with a half and anti-commutes with
the remaining half of the $4^n$ operators of the $n$-qubit Lie
algebra~\cite{Sarkar:2019/arXiv/1909.08123}. Thus, ``on average'' the
commutator $[\hat H, \hat P]$ will contain only half the terms,
$\approx M/2$. In total, we have $\sim \frac{3}{2}M$ terms in
$\hat{H}_{d}^{(1)}$ after the dressing with one generator, or
$\sim (3/2)^{N_g} M$ after a dressing with $N_g$ of them.

\section{Qubit reduction procedure for general operators}
\label{sec:qubit-reduct-proc}

For systems with a small number of qubits, the identification and
removal of stationary qubits provide a noticeable reduction of
computational complexity. Operators that commute with the Hamiltonian
represent exact symmetries. Qubits in such operators can be reduced in
the same manner as those in the Hamiltonian; see
Sec.~\ref{sec:electr-struct-calc}. We derived reduced qubit
expressions for the total electron number operator, $\hat N$, and the
total spin operators $\hat S_z$ and $\hat S^2$. The corresponding full
qubit expressions were derived from the second-quantized counterparts
by applying the chosen fermion-to-qubit transformations (see
Table~\ref{tab:op_prop}).

However, not every operator can be reduced, as its symmetry may not be
compatible with the symmetry of the Hamiltonian. In particular, the
\gls{UCC} ansatz, whose cluster operators are written in a
spin-orbital basis, preserves the electron-number but not the spin
symmetry. Some of $\hat T$ operators do not commute with $\hat S^2$.
Such operators couple different spin sub-blocks of the Hamiltonian,
and their qubit expressions have operators other than $\hat 1$ or
$\hat z$ at the position of the stationary qubits. To find reducible
combinations, we derived spin-adapted cluster amplitudes following a
general scheme given by Eqs.~13.7.2 and 13.7.2 of
Ref.~\citenum{Helgaker:2000}. In particular, we solved equations
\begin{align}
  \label{eq:P_sf}
  [\hat S_{\pm}, \hat P_k] & = 0, \\
  [\hat S_z, \hat  P_k] & = 0, \quad k = 1\,\text{(singles)},\ 
                          2\,\text{(doubles)}, \ldots
\end{align}
for the qubit expressions of $\hat P_k$, $\hat S_z$, and
$\hat S_{\pm}$ (spin raising/lowering) operators. The solutions are
spin-free cluster amplitudes that have stationary qubits at the same
positions as the Hamiltonian, and thus can be reduced by the same
procedure.

\bibliography{iqcc-loc,qcomp-snap,books,programs,ekt-snap,databases,gp}

\end{document}